\documentclass[10pt]{article}
\pdfoutput=0

\usepackage{amsmath}
\usepackage{amssymb}
\usepackage{graphicx}

\usepackage{cite}

\usepackage{color} 

\usepackage{setspace} 
\usepackage[labelfont=bf,labelsep=period,justification=raggedright]{caption}

\makeatletter
\renewcommand{\@biblabel}[1]{\quad#1.}
\makeatother

\date{}

\newcommand{\ie}{i.e., }

\newcommand{\unit}[1] {\,\mathrm{#1}}
\newcommand{\chem}[1] {\ensuremath{\mathrm{#1}}}

\begin{document}

\begin{flushleft}
{\Large
\textbf{Mapping Enzymatic Catalysis using the Effective Fragment Molecular Orbital Method: Towards all ab initio Biochemistry}
}
\\
Casper Steinmann$^{1}$, 
Dmitri G. Fedorov$^{2}$, 
Jan H. Jensen$^{1,\ast}$
\\
\bf{1} Department of Chemistry, University of Copenhagen, Universitetsparken 5, DK-2100 Copenhagen, Denmark
\\
\bf{2} NRI, National Institute of Advanced Industrial Science and Technology (AIST), 1-1-1 Umezono, Tsukuba, Ibaraki 305-8568, Japan
\\
$\ast$ corresponding author, E-mail: jhjensen@chem.ku.dk
\end{flushleft}

\begin{abstract}
We extend the Effective Fragment Molecular Orbital (EFMO) method to the frozen domain approach where only the geometry of an active part is optimized, while the many-body polarization effects are considered for the whole system. The new approach efficiently mapped out the entire reaction path of chorismate mutase in less than four days using 80 cores on 20 nodes, where the whole system containing 2398 atoms is treated in the ab initio fashion without using any force fields. The reaction path is constructed automatically with the only assumption of defining the reaction coordinate a priori. We determine the reaction barrier of chorismate mutase to be {\color{black}$18.3\pm 3.5$ kcal mol$^{-1}$ for MP2/cc-pVDZ and $19.3\pm 3.6$ for MP2/cc-pVTZ in an ONIOM approach using EFMO-RHF/6-31G(d)} for the high and low layers, respectively.
\end{abstract}

\section*{Introduction}
Fragment-based quantum mechanical methods \cite{Otto1975,Gao1997,GEBF,EEMB,SMF,KEM,Ryde2010,MTA,ELG,DC,XO-ONIOM,MFCC-D} are becoming increasingly popular\cite{gordon2012fragmentation}, and have been used to describe a very diverse set of molecular properties for large systems. Although these methods have been applied to refine the energetics of some enzymatic reactions\cite{ishida2006all,nemukhin2012} they are usually not efficient enough to allow for many hundreds of single point calculations needed to map out a reaction path for a system containing thousands of atoms, although geometry optimizations of large systems can be performed for systems consisting of several hundreds of atoms \cite{XO-ONIOM,MTA,ELG,FMOopt,FMOMP2_PCM,fedorov2011geometry}. In fact, typically applications of fragment-based methods to biochemical systems, for example, to protein-ligand binding \cite{Sawada}, are based on performing a few single point calculations for structures obtained at a lower level of theory (such as with force fields). Although many force fields are well tuned to treat typical proteins, for ligands they can be problematic.

In this work we extend the effective fragment molecular orbital (EFMO) method\cite{steinmann2010effective,steinmann2012effective} 
into the frozen domain (FD) formalism \cite{fedorov2011geometry}, originally developed for the fragment molecular orbital (FMO) method \cite {FMO1,FMOrev1,FMObook,FMOrev2}. For FMO, there is also partial energy gradient method \cite{PEG}.

EFMO is based on dividing a large molecular system into fragments and performing ab initio calculations of fragments and their pairs, and combining their energies in the energy of the whole system (see more below). In the FD approach we employ here, one defines an active region associated with the active site, and the cost of a geometry optimization is then essentially given by the cost associated with the active region.  

However, unlike the quantum-mechanical / molecular mechanical (QM/MM) method \cite{MD-rev4} with non-polarizable force fields, the polarization of the whole
system is accounted for in FMO and EFMO methods: in the former via the explicit polarizing potential and in the latter via
fragment polarizabilities. Another important difference between EFMO and QM/MM is that the former does not involve force fields, and 
the need to elaborately determine parameters for ligands does not exist in EFMO. Also, in EFMO all fragments are treated with quantum mechanics,
and the problem of the active site size \cite{Ryde} does not arise.

The paper is organized as follows: First, we derive the EFMO energy and gradient expressions for the frozen domain approach, when some part of the system is frozen during the geometry optimization. Secondly, we predict the reaction barrier of barrier of the conversion of chorismate to prephenate (Figure~\ref{fig:chorismate}) in chorismate mutase. The reaction has been studied previously using conventional QM/MM techniques \cite{lyne1995insights,davidson1996mechanism,hall2000aspects,marti2001transition,worthington2001md,lee2002reaction,lee2003exploring,ranaghan2003insights,ranaghan2004transition,friesner2005ab,crespo2005multiple,claeyssens2011analysis}. The EFMO method is similar in spirit to QM/MM in using a cheap model for the less important part of the system and the mapping is accomplished with a reasonable amount of computational resources (four days per reaction path using 80 CPU cores). Finally we summarize our results and discuss future directions.



\section*{Background and Theory}
The EFMO energy of a system of $N$ fragments (monomers) is
\begin{equation}
E^\mathrm{EFMO} = \sum_I^N E_I^0 + \sum_{IJ}^{R_{I,J}\leq R_\mathrm{resdim}} \left( \Delta E_{IJ}^0 - E_{IJ}^\mathrm{POL} \right) + \sum_{IJ}^{R_{I,J} > R_\mathrm{resdim}} E_{IJ}^\mathrm{ES} + E_\mathrm{tot}^\mathrm{POL}
\label{eqn:efmoenergy}
\end{equation}
where $E_I^0$ is the gas phase energy of monomer $I$. The second sum in equation~\ref{eqn:efmoenergy} is the pairwise correction to the monomer energy and only applies for pairs of fragments (dimers) separated by an interfragment distance $R_{I,J}$ (defined previously \cite{steinmann2010effective}) less than a threshold $R_\mathrm{resdim}$. The correction for dimer $IJ$ is
\begin{equation}
\Delta E_{IJ}^0 = E_{IJ}^0 - E^0_I - E^0_J.
\end{equation}
$E_{IJ}^\mathrm{POL}$ and $E_\mathrm{tot}^\mathrm{POL}$ are the classical pair polarization energy of dimer $IJ$ and the classical total polarization energy, respectively. Both energies are evaluated using the induced dipole model\cite{day1996effective,gordon2001effective} based on distributed polarizabilities\cite{minikis2001accurate}. The final sum over $E_{IJ}^\mathrm{ES}$ is the classical electrostatic interaction energy and applies only to dimers separated by a distance greater than $R_\mathrm{resdim}$. These energies are evaluated using atom-centered multipole moments through quadrupoles\cite{stone1981distributed}. The multipole moments and distributed polarizabilities are computed on the fly for each fragment\cite{steinmann2010effective,steinmann2012effective}.


In cases where only part of a molecular system is to be optimized by minimizing the energy, equation~\ref{eqn:efmoenergy} can be rewritten, resulting in a method conceptually overlapping with QM/MM in using a cheap model for the less important part of the system. Consider a system $S$ (Figure~\ref{fig:active}) where we wish to optimize the positions of atoms in region $A$, while keeping the atoms in region $b$ and $F$ frozen (the difference between $b$ and $F$ will be discussed below). With this definition, we rewrite the EFMO energy as
\begin{equation} \label{eqn:energyregion}
E^\mathrm{EFMO} = E^0_F + E^0_b + E^0_A + E^0_{F/b} + E^0_{F/A} +E^0_{A/b} + E_\mathrm{tot}^\mathrm{POL},
\end{equation}
where $E^0_A$ is the internal energy of region $A$. Region $A$ is made of fragments contaning atoms whose position is optimized,
and $A$ can also have some frozen atoms
\begin{equation}
E^0_A = \sum_{I\in A}^N E_I^0 + \sum_{{I,J\in A}}^{R_{I,J}\leq R_\mathrm{resdim}} \left( \Delta E_{IJ}^0 - E_{IJ}^\mathrm{POL} \right) + \sum_{{I,J\in A}}^{R_{I,J} > R_\mathrm{resdim}} E_{IJ}^\mathrm{ES}.
\end{equation}
Similarly, $E^0_b$ is the internal energy of $b$
\begin{equation} \label{eqn:regionbenergy}
E^0_b = \sum_{I\in b}^N E_I^0 + \sum_{{I,J\in b}}^{R_{I,J}\leq R_\mathrm{resdim}} \left( \Delta E_{IJ}^0 - E_{IJ}^\mathrm{POL} \right) + \sum_{{I,J\in b}}^{R_{I,J} > R_\mathrm{resdim}} E_{IJ}^\mathrm{ES},
\end{equation}
Region $A$ is surrounded by a buffer $b$, because fragment pairs computed with QM containing one fragment outside of $A$ (i.e., in $b$) can still contribute to the total energy gradient (see below). On the other hand, fragment pairs with one fragment in $F$ can also
contribute to the total gradient, but they are computed using a simple classical expression rather than with QM. 
Note that the relation between the notation used in FMO/FD and that we use here is as follows: $A,F$ and $S$ are the same. The buffer region $B$ includes $A$, but $b$ does not, i.e., $A$ and $b$ share no atoms. Formally, $A$ and $b$ are always treated at the same level of theory by assigning fragments to the same layer.

\begin{figure}[!ht]
\begin{center}
\includegraphics[width=5in]{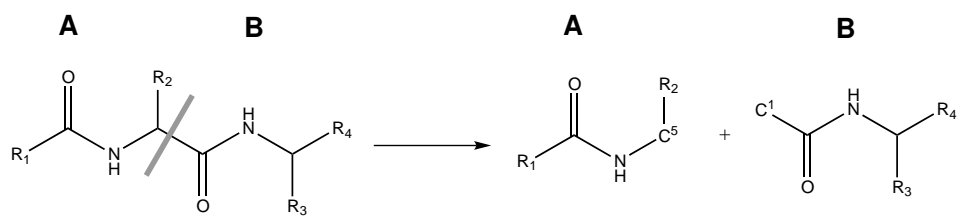}
\end{center}
\caption{
{\bf Cross region fragmentation}. The fragmentation procedure shares an atom {\color{black}(here C$^1$ and C$^5$ is the shared atom)} between two neighboring and covalently bonded fragments. Even though these fragments are in separate regions, they still share an atom across that region as illustrated.}
\label{fig:bondregion}
\end{figure}

{\color{black}In the EFMO method, covalent bonds between fragments are not cut. Instead, electrons from a bond connecting two fragments are placed entirely to one of the fragments. The electrons of the fragments are kept in place by using frozen orbitals across the bond. \cite{fedorov2008covalent,fedorov2009analytic,steinmann2012effective} Fragments connected by a covalent bond} share atoms (Figure~\ref{fig:bondregion}) through the bonding region so it is possible that one side changes the wave function of the bonding region\cite{steinmann2012effective}. It is therefore necessary to re-evaluate the internal \emph{ab initio} energy of region $b$ for each new geometry step. 

The internal geometries of fragments in region $F$ are completely frozen so the internal energy is constant and is therefore neglected
\begin{equation}
\label{eqn:eintfrozen} E^0_F = 0.
\end{equation}
However, it is still necessary to compute the multipole moments and polarizability tensors (and therefore the wave function) of the fragments in $F$ once at the beginning of a geometry-optimization to evaluate $E_\mathrm{tot}^\mathrm{POL}$ in equation~\ref{eqn:energyregion} as well as some inter-region interaction energies defined as
\begin{align}
\label{eqn:crossba} E^0_{b/A} &= \sum_{\substack{I\in b\\ J\in A}}^{R_{I,J}\leq R_\mathrm{resdim}} \left(\Delta E_{IJ}^0 - E_{IJ}^\mathrm{POL} \right) + \sum_{\substack{I\in b\\ J\in A}}^{R_{I,J} > R_\mathrm{resdim}} E_{IJ}^\mathrm{ES}, \\
\label{eqn:crossfa} E^0_{F/A} &= \sum_{\substack{I\in A\\ J\in F}} E_{IJ}^\mathrm{ES}, \\
\label{eqn:crossfb} E^0_{F/b} &= 0.
\end{align}
Equation~\ref{eqn:crossfa} assumes that $b$ is chosen so that fragments in $A$ and $F$ are sufficiently separated (\ie $R_{I,J} > R_\mathrm{resdim}$) so the interaction is evaluated classically. If all atoms in region $b$ are frozen, then $E^0_{F/b}$ is constant and can be neglected. However, this assumes that the positoins of all atoms at both sides of the bonds connecting fragments are frozen.

The final expression for the EFMO frozen domain (EFMO/FD) energy is
\begin{equation} \label{eqn:efmofinal}
E^{\mathrm{EFMO}} = E^0_b + E^0_A + E^0_{b/A} + \sum_{\substack{I\in A\\ J\in F}}^{R_{I,J} > R_\mathrm{resdim}} E_{IJ}^\mathrm{ES} + E_\mathrm{tot}^\mathrm{POL}.
\end{equation}
Finally, we note that due to the frozen geometry of $b$ we can further gain a speedup by not evaluating dimers in $b$ (cross terms between $A$ and $b$ are handled explicitly according to equation~\ref{eqn:crossba}) since they do not contribute to the energy or gradient of $A$. This corresponds to the frozen domain with dimers (EFMO/FDD), and equation~\ref{eqn:regionbenergy} becomes 
\begin{equation}
\label{eqn:approxregionb}
E^0_b = \sum_{I\in b}^N E_I^0.
\end{equation}
The gradient of each region is
\begin{align}
\frac{\partial E^{\mathrm{EFMO}}}{\partial x_A} &= \frac{\partial E^0_A}{\partial x_A} + \frac{\partial E^0_{A/b}}{\partial x_A} + \frac{\partial E^0_{A/F}}{\partial x_A} + \frac{\partial E_\mathrm{tot}^\mathrm{POL}}{\partial x_A}, \\
\label{eqn:gradregionb}\frac{\partial E^{\mathrm{EFMO}}}{\partial x_b} &= 0, \\
\frac{\partial E^{\mathrm{EFMO}}}{\partial x_F} &= 0,
\end{align}
and the details of their evaluation has been discussed previously\cite{steinmann2010effective,steinmann2012effective}. Equation~\ref{eqn:gradregionb} does not apply to non-frozen atoms shared with region $A$.

The frozen domain formulation of EFMO was implemented in GAMESS\cite{schmidt1993general} and parallelized using the generalized distributed data interface\cite{FMO3,fedorov2004new}.

\section*{Computational Details}

\subsection*{Preparation of the Enzyme Model}
\begin{figure}[!ht]
\begin{center}
\includegraphics[width=5in]{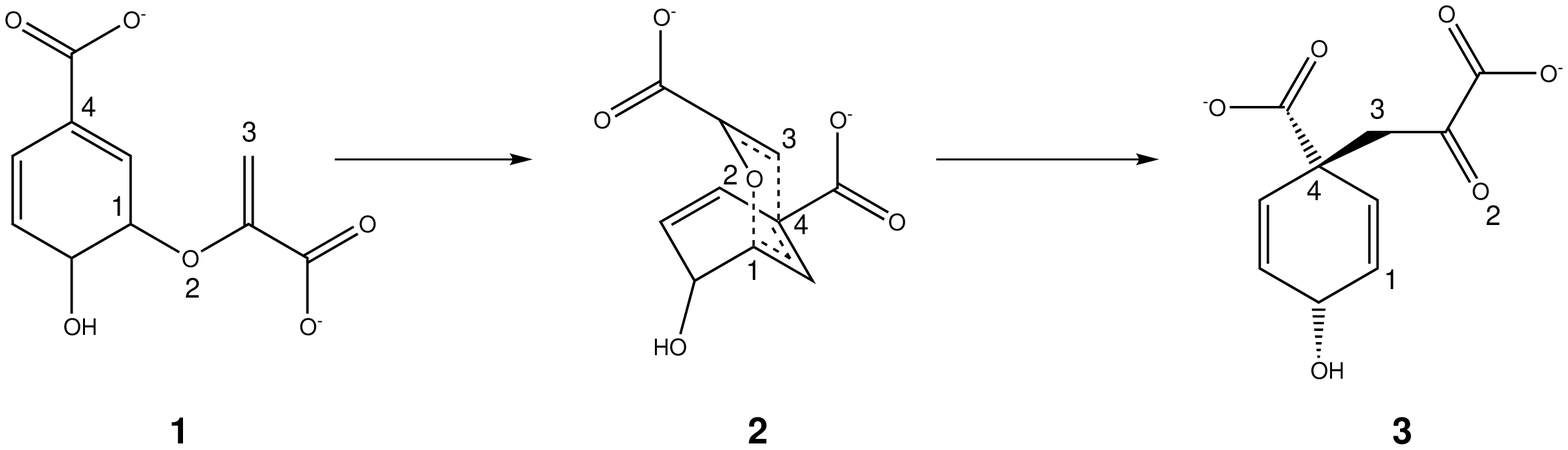}
\end{center}
\caption{
{\bf Conversion of chorismate to prephenate through its transition state}. Atoms of interest are marked with numbers one trough four.}
\label{fig:chorismate}
\end{figure}

We followed the strategy by Claeyssens \emph{et~al.}\cite{claeyssens2011analysis} The structure of chorismate mutase (PDB: 2CHT) solved by Chook \emph{et~al}.\cite{chook1993crystal} was used as a starting point. Chains A, B and C were extracted using PyMOL\cite{schrodinger2010pymol} and subsequently protonated with PDB2PQR\cite{dolinsky2004pdb2pqr,dolinsky2007pdb2pqr} and PROPKA\cite{li2005very} at $\mathrm{pH} = 7$. {\color{black}The protonation state of all residues can be found in Table~S1.} The inhibitor between chain A and C was replaced with chorismate in the reactant state (\textbf{1}, Figure~\ref{fig:chorismate}) modeled in Avogadro\cite{avogadro,avogadro2012paper}.

The entire complex (chorismate mutase and chorismate) was solvated in water (TIP3P\cite{jorgensen1983comparison}) using GROMACS. \cite{van2005gromacs,hess2008gromacs} {\color{black} To neutralize the system 11 Na$^{+}$ counter ions were added.} The protein and counter ions were treated with the CHARMM27\cite{kerell2000development,brooks2009charmm} force field in GROMACS. Force-field parameters for chorismate were generated using the SwissParam\cite{zoete2011swissparam} tool. To equilibrate the complex a $100$ ps NVT run at $T=300\unit{K}$ was followed by a $100\unit{ps}$ NPT run at $P=1\unit{bar}$ and $T=300\unit{K}$. The production run was an isothermal-isobaric trajectory run for $10$ ns. A single conformation was randomly selected from the last half of the simulation and energy minimized in GROMACS to a force threshold of $F_\mathrm{max}=300 \unit{KJ}\unit{mol^{-1}}\unit{nm^{-1}}$. During equilibration and final molecular dynamics (MD) simulation, the $\chem{C}3$ and $\chem{C}4$ atoms of chorismate (see Figure~\ref{fig:chorismate}) were harmonically constrained to a distance of 3.3 {\AA} to keep it in the reactant state. Finally, a sphere of 16 {\AA} around the C1 atom of chorismate was extracted in PyMOL and hydrogens were added to correct the valency where the backbone was cut. The final model  contains a total of 2398 atoms.

\subsection*{Mapping the Reaction Path}
To map out the reaction path, we define the reaction coordinate similarly to Claeyssens \emph{et~al.}\cite{claeyssens2011analysis} as the difference in bond length between the breaking O2-C1 bond and the forming C4-C3 bond in chorismate (see also Figure~\ref{fig:chorismate}), i.e.
\begin{equation}
\label{eqn:reactioncoordinate}
R = R_\mathrm{21} -R_\mathrm{43}.
\end{equation}
The conversion of chorismate ($R=-2.0$~{\AA}, $R_{21} = 1.4$~{\AA}, $R_{43} =-3.4$~{\AA}) to prephenate ($R=1.9$ {\AA}, $R_{21} = 3.3$~{\AA}, $R_{43} =1.4$~{\AA}) in the enzyme was mapped by constraining the two bond lengths in equation~\ref{eqn:reactioncoordinate} with a harmonic force constant of 500 kcal mol$^{-1}$ \AA$^{-2}$ in steps of $0.1$~{\AA}. For each step, all atoms in the active region ($A$) were minimized to a threshold on the gradient of $5.0\cdot10^{-4}$ Hartree Bohr$^{-1}$ (OPTTOL=5.0e-4 in \$STATPT). For the enzyme calculations we used EFMO-RHF and FMO2-RHF with the frozen domain approximation presented above. 

\begin{figure}[!ht]
\begin{center}
\includegraphics[width=5cm]{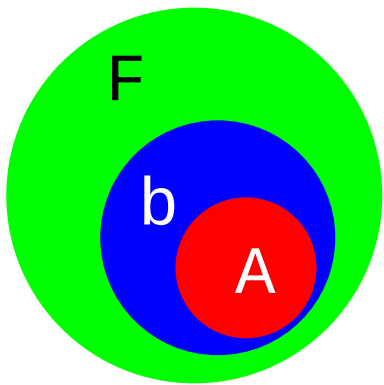}
\end{center}
\caption{
{\bf Definition of a system with active, buffer and frozen regions in frozen domain EFMO.}}
\label{fig:active}
\end{figure}

\begin{figure}[!ht]
\begin{center}
\includegraphics[width=5in]{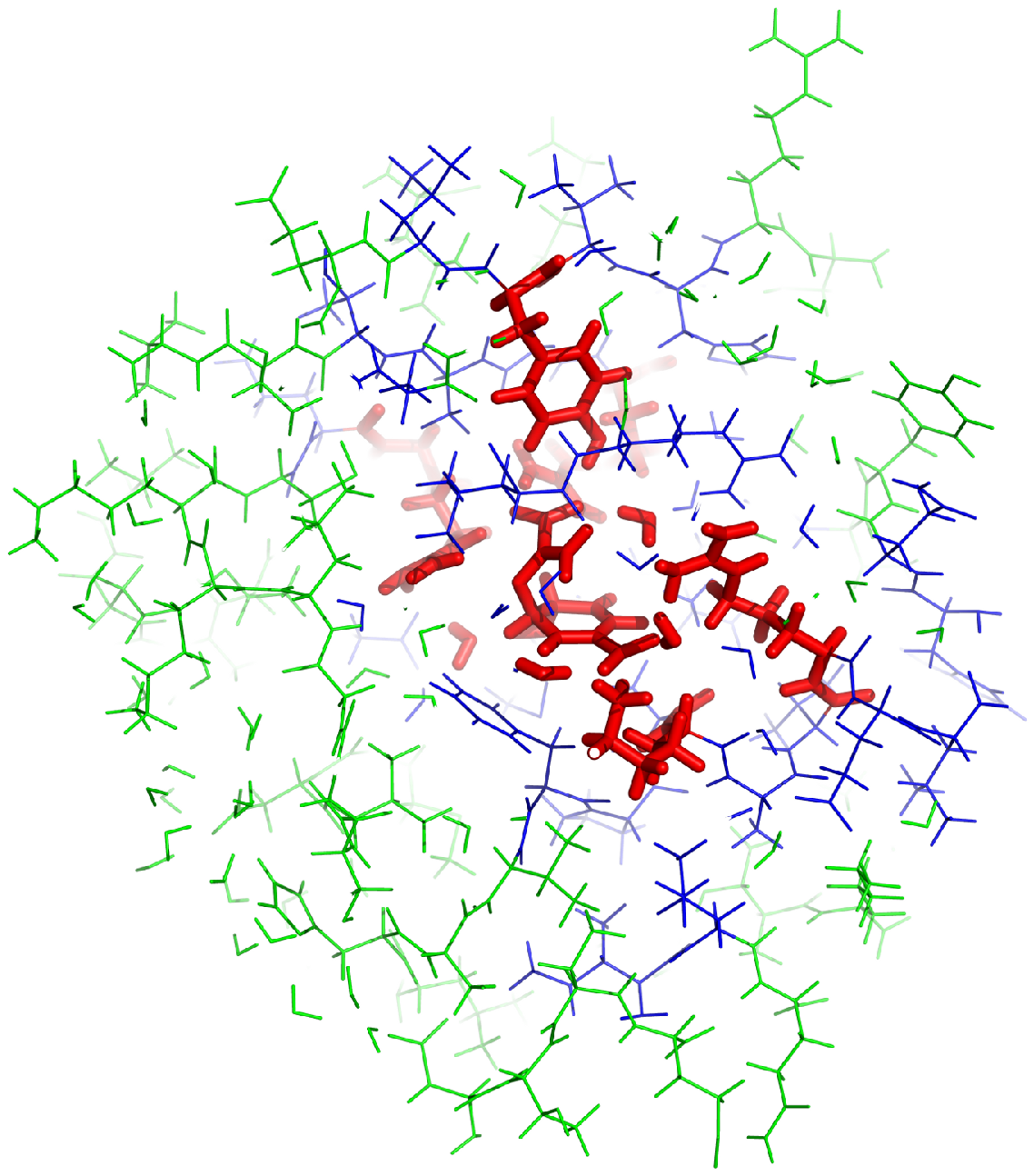}
\end{center}
\caption{
{\bf EFMO:S model of chorismate mutase used in this study}. The entire model contains 2398 atoms. There are 1341 atoms in green belonging to the frozen region ($F$), 928 atoms in blue belonging to the buffer region ($b$) and 129 atoms in red belonging to the active region ($A$).}
\label{fig:modelfull}
\end{figure}

\begin{figure}[!ht]
\begin{center}
\includegraphics[width=5in]{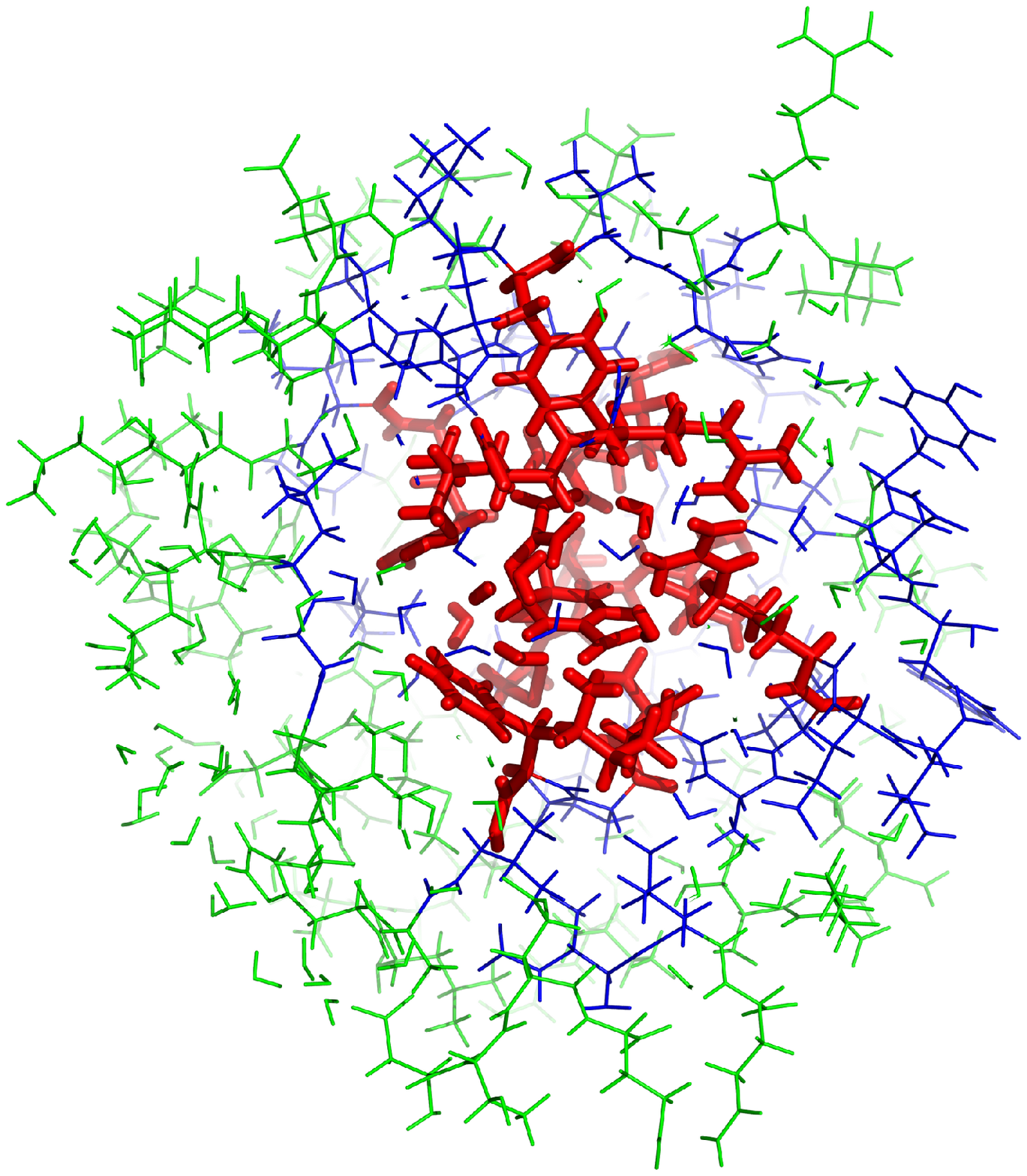}
\end{center}
\caption{
{\bf EFMO:L model of chorismate mutase used in this study}. The entire model contains 2398 atoms. There are 1006 atoms in green belonging to the frozen region ($F$), 1151 atoms in blue belonging to the buffer region ($b$) and 241 atoms in red belonging to the active region ($A$).}
\label{fig:modelfull_l}
\end{figure}

We used two different sizes for the active region small: (EFMO:$\mathbf{S}$, Figure~\ref{fig:modelfull}) and large (EFMO:$\mathbf{L}$, Figure~\ref{fig:modelfull_l}). The active region (colored red in Figure~\ref{fig:active}) is defined as all fragments with a minimum distance $R_\mathrm{active}$ from any atom in chorismate (EFMO:$\mathbf{S}:R_\mathrm{active}=2.0$ {\AA}, EFMO:$\mathbf{L}:R_\mathrm{active}=3.0$ {\AA}). In EFMO:$\mathbf{S}$ the active region consists of chorismate, 4 residues and 5 water molecules, while the active region in EFMO:$\mathbf{L}$ consists of chorismate, 11 residues and 4 water molecules. The buffer region (blue in Figure~\ref{fig:active}) is defined as all fragments within $2.5$ {\AA} of the active region for both EFMO:$\mathbf{S}$ and EFMO:$\mathbf{L}$. The rest of the system is frozen. To prepare the input files we used FragIt\cite{steinmann2012fragit}, which automatically divides the system into fragments; in this work we used the
fragment size of one amino acid residue or water molecule per fragment.

In order to refine the energetics, for each minimized step on the reaction path we performed two-layer ONIOM\cite{svensson1996oniom,dapprich1999new} calculations
\begin{equation}
E^\mathrm{high}_\mathrm{real} \approx E^\mathrm{low}_\mathrm{real} + E^\mathrm{high}_\mathrm{model} - E^\mathrm{low}_\mathrm{model},
\end{equation}
where $E^\mathrm{low}_\mathrm{real} = E^\mathrm{EFMO}$ according to equation~\ref{eqn:energyregion}. This can be considered a special
case of the more general multicenter ONIOM based on FMO \cite{ONIOM-FMO}, using EFMO instead of FMO. The high level model system is chorismate in the gas-phase calculated using B3LYP\cite{becke1993new,stephens1994ab,hertwig1997parameterization} (DFTTYP=B3LYP in \$CONTRL) or MP2 (MPLEVL=2 in \$CONTRL) with either 6-31G(d) or the cc-pVDZ{\color{black}, cc-pVTZ and cc-pVQZ basis sets} by Dunning \cite{dunning1989gaussian}.

We also carried out multilayer EFMO and FMO\cite{fedorov2005multilayer} single-point calculations where region $F$ is described by RHF/6-31G(d) and $b$ and $A$ (for EFMO) or $B$ ($B=A \cup b$ for FMO \cite{fedorov2011geometry}) is calculated using MP2/6-31G(d).


The FDD approximation in equation~\ref{eqn:approxregionb} is enabled by specifying MODFD=3 in \$FMO, similarly to the frozen domain approach in FMO\cite{fedorov2011geometry}. All calculations had spherical contaminants removed from the basis set (ISPHER=1 in \$CONTRL).

\subsection*{Obtaining the Activation Enthalpy}
\begin{figure}[!ht]
\begin{center}
\includegraphics[width=5in]{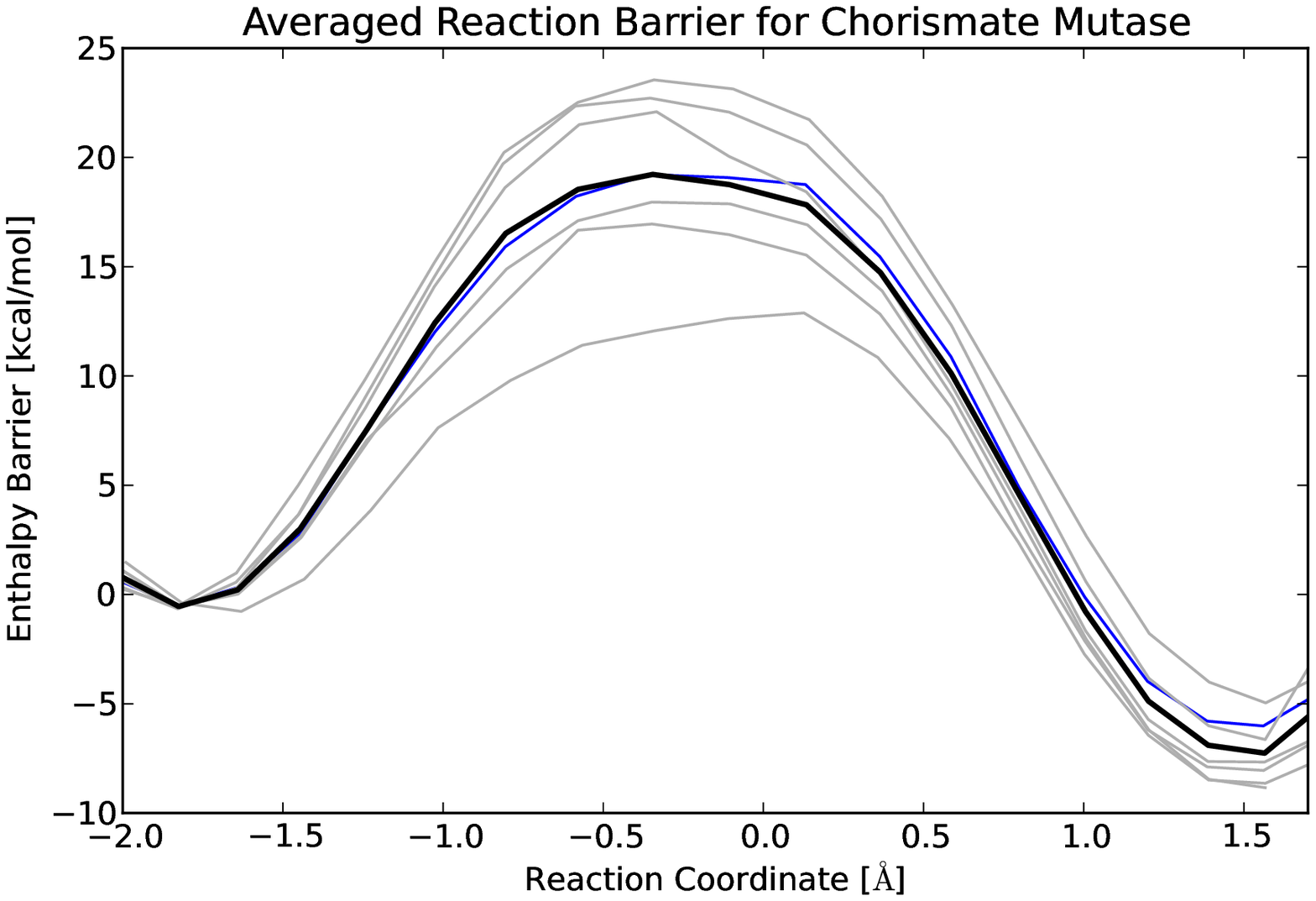}
\end{center}
\caption{
{\bf Reaction Enthalpy Profile for chorismate mutase.} The 7 profiles are calculated with ONIOM at the MP2/cc-pVDZ:EFMO-RHF/6-31G(d) level of theory. The black line is the average reaction energy, grey lines are individual reaction paths. The blue line is the reaction path discussed in detail, in the results section.}
\label{fig:average}
\end{figure}

The activation enthalpy is obtained in two different ways by calculating averages of $M$ adiabatic reaction pathways. The starting points of the $M$ pathways were randomly extracted from the MD simulation, followed by the reaction path mapping procedure described above for each pathway individually. One way to obtain the activation enthalpy averages the barriers from each individual adiabatic reaction path \cite{claeyssens2006high}
\begin{equation}
\label{eqn:enthalpy1}
\Delta H_1^\ddagger = \frac{1}{M} \sum_{i=1}^M \left(E_{\mathrm{TS},i} - E_{\mathrm{R},i}\right) -1.6\,\mathrm{kcal\,mol}^{-1}.
\end{equation}
Here $M$ is the number of reaction paths ($M = 7$, Figure~\ref{fig:average}) $E_{\mathrm{TS},i}$ is the highest energy on the adiabatic reaction path while $E_{\mathrm{R},i}$ is the lowest energy with a negative reaction coordinate. 1.6 kcal mol$^{-1}$ corrects for the change in zero point energy and thermal contributions\cite{claeyssens2006high}.

The other way of estimating the activation enthalpy is \cite{ranaghan2004transition}:
\begin{equation}
\label{eqn:enthalpy2}
\Delta H_2^\ddagger = \langle E_{\mathrm{TS}}\rangle - \langle E_\mathrm{R}\rangle - 1.6\,\mathrm{kcal\,mol}^{-1}.
\end{equation}
Here $\langle E_{\mathrm{TS}}\rangle$ and $\langle E_\mathrm{R}\rangle$ are, respectively, the highest energy and lowest energy with a negative reaction coordinate on the averaged adiabatic path (bold line in Figure~\ref{fig:average}). The brackets here mean averaging over 7 reaction paths; and the difference of Eqs \ref{eqn:enthalpy1} and \ref{eqn:enthalpy2} arises because of the noncommutativity of the sum and the min/max operation over coordinates: in Eq \ref{eqn:enthalpy1} we found a minimum and a maximum for each curve separately, and averaged the results, but in Eq \ref{eqn:enthalpy2} we first averaged and then found the extrema. As discussed below, the two reaction enthalpies are within 0.2 kcal/mol, which indicates that the TS occurs at roughly the same value of the reaction coordinate for most paths.

\section*{Results and Discussion}

\subsection*{Effects of Methodology, Region Sizes and Approximations}
\label{sect1}
Reaction barriers obtained in the enzyme using harmonic constraints are plotted on Figure~\ref{fig:rhfbarrier} and listed in Table~\ref{tbl:rhfbarriers} for different settings of region sizes and approximations. All calculated reaction barriers are within 0.5 kcal mol$^{-1}$ from each other when going from the reactant ($R_R$) to the proposed transition ($R_{TS}$) state where the reaction barriers for the TSs are around 46 kcal mol$^{-1}$. The same is true when going to the product $R_P$. Only the large model (EFMO:$\mathbf{L}$) shows a difference in energy near the product ($R_P$) with a lowering of the relative energy by 4 kcal mol$^{-1}$ compared to the other settings. 

The reaction coordinates are also similar for the small systems ($R_P=1.41$ \AA, except for $R_\mathrm{resdim}=2.0$ which is $R_P=1.56$ \AA) with some minor kinks on the energy surface from optimization of the structures without constraints at $R_P$. The EFMO:\textbf{L} model has a different reaction coordinate for the product ($R_P=1.57$ \AA) and also a shifted reaction coordinate for the transition state $R_{TS}=-0.12$ \AA\, which we can attribute to a better description of more separated pairs in the active region but more importantly that around the TS, the energy surface is very flat. Interestingly, using FMO2 shows no significant change in either reaction barriers or reaction coordinates for the reactant, transition state or product which differ from EFMO:\textbf{S} by 0.02 \AA, 0.03 \AA\, and 0.01 \AA\, respectively. Timings are discussed below.

Previous work by Ranaghan \emph{et~al.}\cite{ranaghan2003insights,ranaghan2004transition} obtained an RHF barrier of $36.6$ kcal mol$^{-1}$ which is 10 kcal/mol lower than what we obtained. Also, they observed that the transition state happened earlier at $R_{TS} = -0.3$ \AA. The difference in reaction barrier from our findings is attributed to a poorer enzyme structure and other snapshots do yield similar or better reaction barriers (see below). Furthermore, the same study by Ranaghan et al. found that the reaction is indeed exothermic with a reaction energy of around $-30$ kcal mol$^{-1}$ at the RHF/6-31G(d) level of theory. We expect this difference from our results to arise from the fact the study by Ranaghan et al. used a fully flexible model for both the substrate and the enzyme where the entire protein is free to adjust contrary to our model where we have chosen active fragments and atoms in a uniform sphere around a central fragment. This is perhaps not the best solution if one includes too few fragments (which lowers the computational cost) due to fragments in the buffer region are unable to move and cause steric clashes. The lowering of the energy for EFMO:\textbf{L} suggests this.


\subsection*{Refined Reaction Energetics}
\label{sect2}
For the smallest EFMO:\textbf{S} system ONIOM results are presented on Figure~\ref{fig:oniombarrier} and in Table~\ref{tbl:oniombarriers} for various levels of theory. By calculating the MP2/cc-pVDZ:EFMO-RHF/6-31G(d) energy using ONIOM we obtain a 19.8 kcal mol$^{-1}$ potential energy barrier. Furthermore, the reaction energy is lowered from $-1.3$ kcal mol$^{-1}$ to $-5.5$ kcal mol$^{-1}$. {\color{black}Increasing the basis set size through cc-pVTZ and cc-pVQZ reduces the barrier to 21.8 kcal mol$^{-1}$ and 21.7 kcal mol$^{-1}$, respectively and the reaction energy is -1.1 kcal mol$^{-1}$ and 0.8 kcal mol$^{-1}$.} Using the smaller 6-31G(d) basis set with MP2, the reaction barrier is 22.2 kcal mol$^{-1}$ and reaction energy is $-3.2$ kcal mol$^{-1}$. The B3LYP results are improvements for the TS only reducing the barrier to $23.8$ kcal mol$^{-1}$ for B3LYP/cc-pVDZ:EFMO-RHF/6-31G(d). The same is not true for the product where the energy is increased by about $3$ kcal mol$^{-1}$. For the other systems treated using EFMO-RHF/6-31G(d) discussed in the previous section ONIOM corrected results {\color{black}at the MP2 or B3LYP level of theory using a cc-pVDZ basis set} are listed in tables~S2 to S5 and show differences from the above by less than $1$ kcal mol$^{-1}$, again the reaction coordinates changes slightly depending on the system. {\color{black}The effect of including correlation effects by means of MP2 and systematically larger basis sets is that the potential energy barrier for the reaction rises as more correlation effects are included, the same is true for the overall reaction energy.}

The results presented here for MP2 are in line with what has been observed previously by Ranaghan \emph{et~al.}\cite{ranaghan2004transition} and Claeyssens \emph{et~al.}\cite{claeyssens2011analysis}. Overall, the reaction barrier is reduced to roughly half of the RHF barrier and the observed coordinates for the reaction shift slightly. We do note that this study and the study by Ranaghan \emph{et~al.} use ONIOM style energy corrections for the correlation and not geometry optimizations done at a correlated level. Overall, we observe that the predicted reaction coordinate for the approximate transition state in the conversion of chorismate to prephenate happens around $0.2$ {\AA} later than in those studies.

The results for the multilayer single points along the energy surface are presented in Table~\ref{tbl:efmomp2barrier}. The barrier calculated at the EFMO-RHF:MP2/6-31G(d) level of theory is predicted to be $27.6$ kcal mol$^{-1}$ which is {\color{black}5.4} kcal mol$^{-1}$ higher than the ONIOM barrier and the reaction coordinates are shifted for both the reactant and the TS from $R_{R}=-1.95$ {\AA} ~to $R_{R}=-1.64$ {\AA} ~and $R_{TS}=-0.36$ {\AA} ~to $R_{TS}=-0.11$ {\AA}. Similar results are obtained at the FMO2-RHF:MP2/6-31G(d) level of theory. {\color{black}The difference from the ONIOM corrected values in table~\ref{tbl:efmomp2barrier} is likely due to the inclusion of dispersion effects between the chorismate and the enzyme which is apparently weaker at the transition state compared to the reactant state.}

\subsection*{Ensemble Averaging}
\begin{figure}[!ht]
\begin{center}
\includegraphics[width=5in]{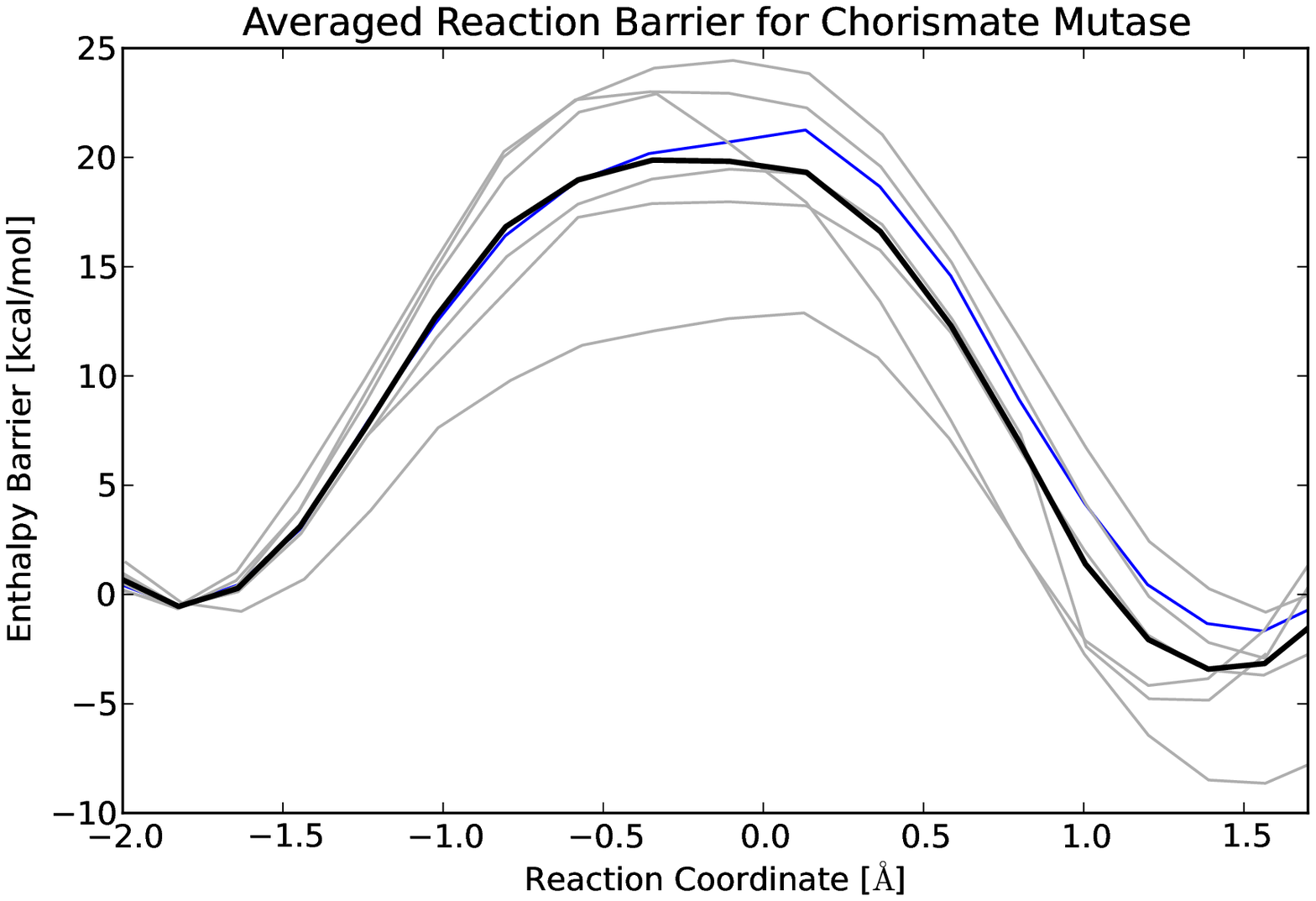}
\end{center}
\caption{
{\color{black}{\bf Reaction Enthalpy Profile for chorismate mutase.} The 7 profiles are calculated with ONIOM at the MP2/cc-pVTZ:EFMO-RHF/6-31G(d) level of theory. The black line is the average reaction energy, grey lines are individual reaction paths. The blue line is the reaction path discussed in detail, in the results section.}}
\label{fig:average_cct}
\end{figure}

\begin{figure}[!ht]
\begin{center}
\includegraphics[width=5in]{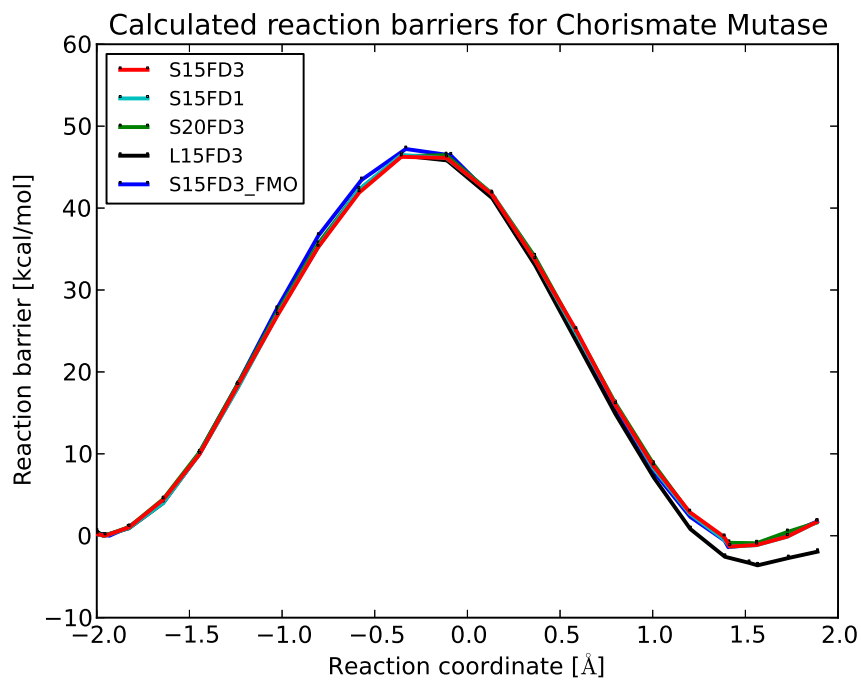}
\end{center}
\caption{
{\bf EFMO-RHF/6-31G(d) barrier for chorismate mutase}. S15FD3 and S15FD3\_FMO are EFMO:\textbf{S} and FMO:\textbf{S}, respectively, both with $R_{\mathrm{resdim}}=1.5$, and the dimer approximation in region $b$ (Equation~\ref{eqn:approxregionb}). S15FD1 is similar to S15FD3 but without the dimer approximation in region $b$. S20FD3 is also similar to S15FD3 but with $R_{\mathrm{resdim}}=2.0$, instead. Finally, L15FD3 is EFMO:\textbf{L} with $R_{\mathrm{resdim}}=1.5$, and the dimer approximation (FDD) in region $b$.}
\label{fig:rhfbarrier}
\end{figure}

\begin{figure}[!ht]
\begin{center}
\includegraphics[width=5in]{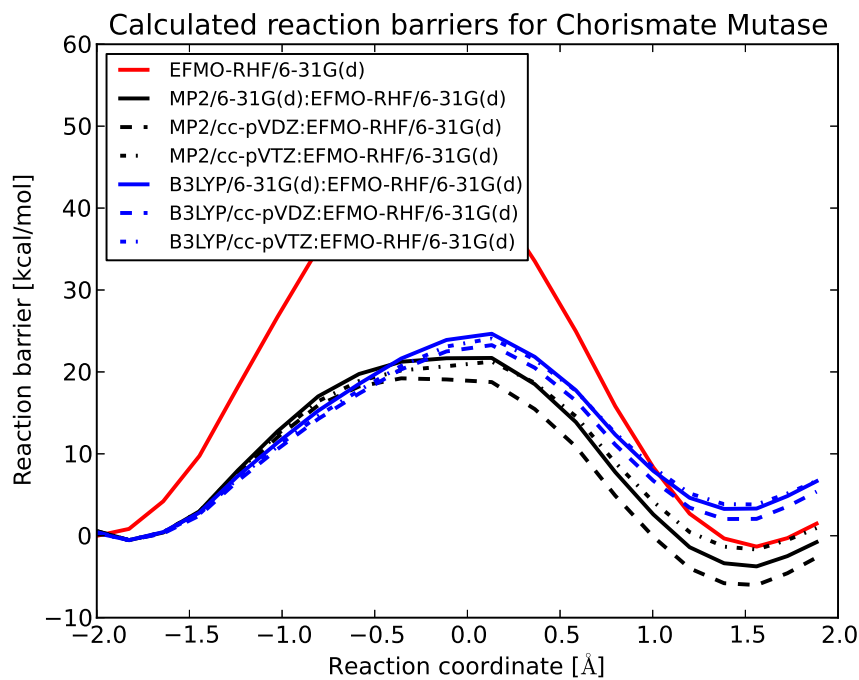}
\end{center}
\caption{
{\bf ONIOM results calculated with various levels of theory for EFMO:\textbf{S}} geometries. The red curve is the EFMO-RHF/6-31G(d) result also presented in Figure~\ref{fig:rhfbarrier}. Blue (B3LYP) and green (MP2) curves are ONIOM results with chorismate calculated in the gas-phase using the 6-31G(d) (solid lines), cc-pVDZ (dashed lines) {\color{black}or cc-pVTZ (dotted line) basis set.}}
\label{fig:oniombarrier}
\end{figure}

In Figure~\ref{fig:average} and Figure~\ref{fig:average_cct} we show 7 adiabatic reaction paths mapped with EFMO-RHF/6-31G(d) starting from 7 MD snapshots; the energetics were refined with ONIOM at the MP2/cc-pVDZ {\color{black}and MP2/cc-pVTZ level}. In EFMO, we used a small active region (EFMO:\textbf{S}) and $R_\mathrm{resdim}=1.5$ and no dimer calculations in region $b$ (S15FD3 in Figure~\ref{fig:rhfbarrier}). Out of the 7 trajectories one is described in detail in the previous sub-section. 

For MP2/cc-pVDZ:EFMO-RHF/6-31G(d) the reaction enthalpies are $\Delta H_1^\ddagger = 18.3 \pm 3.5$ kcal mol$^{-1}$ and $\Delta H_2^\ddagger = 18.2$ kcal mol$^{-1}$ [cf. Equations (\ref{eqn:enthalpy1}) and (\ref{eqn:enthalpy2})], the latter having an uncertainty of the mean of $6.9$ kcal mol$^{-1}$. {\color{black}For MP2/cc-pVTZ:EFMO-RHF/6-31G(d) the reaction enthalpies are $\Delta H_1^\ddagger = 19.3 \pm 3.7$ kcal mol$^{-1}$ and $\Delta H_2^\ddagger = 18.8$ kcal mol$^{-1}$ with an uncertainty of the mean of $7.1$ kcal mol$^{-1}$.} These barriers are ca $5.5$ {\color{black}($6.5$)} kcal mol$^{-1}$ higher than the experimental value of $12.7 \pm 0.4$ kcal mol$^{-1}$ for MP2/cc-pVDZ {\color{black}(MP2/cc-pVTZ)}. For comparison, the activation enthalpy obtained by Claeyssens \emph{et~al.}  \cite{claeyssens2006high,claeyssens2011analysis} ($9.7 \pm  1.8$ kcal mol$^{-1}$) is underestimated by $3.0$ kcal mol$^{-1}$. 

There are several differences between our study and that of Claeyssens \emph{et~al.} that could lead to an overestimation of the barrier height: biasing the MD towards the TS rather than the reactant, a larger enzyme model (7218 vs 2398 atoms), and more conformational freedom when computing the potential energy profile.

With regard to the latter point, while Figure~\ref{fig:rhfbarrier} shows that increasing the active region has a relatively small effect on the barrier this may not be the case for all snapshots.  We did identify one trajectory that failed to produce a meaningful reaction path and is presented in Figure S1. Here, the energy of the barrier becomes unrealistically high because of very little flexibility in the active site and unfortunate placement of Phe57 (located in the buffer region, Figure S2), which hinders the conformational change needed for the successful conversion to prephenate yielding an overall reaction energy of around $+11$ kcal mol$^{-1}$. As noted above, the EFMO:\textbf{L} settings is a possible solution to this as more of the protein in available to move, but as seen from Table~\ref{tbl:rhfbarriers} the computational cost doubles.


\subsection*{Timings}
Using the computationally most efficient method tested here (EFMO:\textbf{S}), $R_\mathrm{resdim}=1.5$, and skipping dimers in the buffer region $b$, an adiabatic reaction path, which requires a total of 467 gradient evaluations, can be computed in four days using 80 CPU cores (20 nodes with 4 cores each) at the RHF/6-31G(d) level of theory. As shown in Table~\ref{tbl:rhfbarriers}, the same calculation using FMO2 requires takes roughly $T^\mathrm{full}_\mathrm{rel}=7.5$ times longer. 

Increasing $R_\mathrm{resdim}$ from 1.5 to 2 has a relatively minor effect of the CPU time (a factor of 1.2), while performing the dimer calculations in the buffer region nearly doubles (1.7) the CPU time.  Similarly, increasing the size of active region from $2.0$ \AA ~to $3.0$ \AA ~around chorismate also nearly doubles (1.8) the CPU time. This is mostly due to the fact that more dimer calculations must be computed, but the optimizations also require more steps (513 gradient evaluations) to converge due to the larger number of degrees of freedom that must be optimized. 

Looking at a single minimization for a specific reaction coordinate $R=-1.79$ {\AA}, the most efficient method takes 4.5 hours. Here, the relative timings $T_\mathrm{rel}$ are all larger than for the full run ($T^\mathrm{full}_\mathrm{rel}$) due to a slight increase in the number of geometry steps (around 25) taken for all but FMO2 which is identical to the reference (22 steps). Thus, the overall cost of performing the FMO2 minimization is 6.7 times as expensive as EFMO.

\section*{Summary and Outlook}
In this paper we have shown that the effective fragment molecular orbital (EFMO) method \cite{steinmann2010effective,steinmann2012effective} can be used to efficiently map out enzymatic reaction paths provided the geometry of a large part of the enzyme and solvent is frozen. In EFMO one defines an active region associated with the active site, and the cost of a geometry optimization is then essentially the cost of running quantum-mechanical calculations of the active domain. This is similar to the cost of QM/MM, if the QM region is the same; the difference is that in EFMO we freeze the coordinates of the rest of the system, whereas in QM/MM they are usually fully relaxed. On the other hand, EFMO does not require parameters and can be better considered an approximation to a full QM calculation rather than a QM/MM approach. 

In this work we used the mapping technique based on running a classical MD simulation, selecting some trajectories, freezing the coordinates of the outside region, and doing constrained geometry optimizations along a chosen reaction coordinate. An alternative to this approach is to run full MD simulation of a chemical reaction using EFMO. This has already been done for many chemical reactions using FMO-MD \cite{fmomd,sato2008does,fmomd-rev} and can be done in future with EFMO.

A potential energy profile for the chorismate to prephenate reaction in chorismate has been computed in 4 days using 80 CPU cores for an RHF/6-31G(d) description of a truncated model of the enzyme containing 2398 atoms. For comparison, a corresponding FMO2 calculation takes about 7.5 times more. The cost of EFMO calculations is mainly determined by the size of the buffer- and active region. Comparing to a QM/MM
with QM region of the same size, EFMO as a nearly linear scaling method, becomes faster than QM if the system size is sufficiently large; especially for correlated methods like MP2 where this cross-over should happen with relatively small sizes.

Our computed conformationally-averaged activation enthalpy is in reasonable agreement to the experimental value, although overestimated by 5.5 kcal/mol.


The energetics of this reaction depends on the level of calculation. We have shown that by using a level better than RHF, for instance, MP2 or DFT, considerably improves the energetics and by using such an appropriate level to also determine the reaction path following the formalism in this work can be used to provide a general and reliable way in future.

EFMO, as one of the fragment-based methods \cite{gordon2012fragmentation}, can be expected to be useful in various biochemical
studies, such as in enzymatic catalysis and protein-ligand binding. It should be noted that in addition to its paramater-free
ab initio based nature, EFMO and FMO also offer chemical insight on the processes by providing subsystem information, such
as the properties of individual fragments (e.g., the polarization energy) as well as the pair interaction energies
between fragments \cite{piedagas,piedapcm}. This can be of considerable use to fragment-based drug discovery \cite{fbdd,gamess-drug}.

\section*{Acknowledgements}
CS and JHJ thank the Danish Center for Scientific Computing at the University of Copenhagen for providing computational resources. DGF thanks the Next Generation Super Computing Project, Nanoscience Program (MEXT, Japan) and Strategic Programs for
Innovative Research (SPIRE, Japan).

\bibliography{efmom}

\newpage
\section{Figures}

\newpage


\newpage

\newpage

\newpage

\newpage

\newpage

\newpage

\newpage

\newpage
\section{Tables}
\begin{table}[!ht]
\caption{\bf{EFMO-RHF and FMO2-RHF results for chorismate mutase}.}
\begin{tabular}{lcccccccll}
Model & $R_{resdim}$ & modfd & $R_R$ & $R_{TS}$ & $R_P$ & $E_{TS-R}$ & $E_{P-R}$ & $T_\mathrm{rel}$ & $T^\mathrm{full}_\mathrm{rel}$\\ \hline
EFMO:\textbf{S} & 1.5 &    3 &  -1.95 &  -0.36 &   1.42 &  46.25 &  -1.32 &    1.0 &    1.0 \\
EFMO:\textbf{S} & 1.5 &    1 &  -1.96 &  -0.36 &   1.42 &  46.49 &  -1.34 &    2.0 &    1.7 \\
EFMO:\textbf{S} & 2.0 &    3 &  -1.96 &  -0.12 &   1.56 &  46.46 &  -0.91 &    1.3 &    1.2 \\
EFMO:\textbf{L} & 1.5 &    3 &  -1.97 &  -0.35 &   1.57 &  46.42 &  -3.61 &    2.1 &    1.8 \\
FMO2:\textbf{S} & 1.5 &    3 &  -1.93 &  -0.33 &   1.41 &  47.21 &  -1.40 &    6.7 &    7.5 \\
\end{tabular}\begin{flushleft}
Reaction barriers of chorismate mutase calculated with different levels of theory. $R_{resdim}$ is unitless. The reaction coordinates for the reactant, transition state and product are $R_\mathrm{R}$, $R_\mathrm{TS}$ and $R_\mathrm{P}$, respectively and given in {\AA}, barrier height of the transition state $E_\mathrm{TS-R}$ and overall reaction energy $E_\mathrm{P-R}$ in kcal/mol. $T_\mathrm{rel}$ are relative timings to EFMO-RHF/6-31G(d) using the EFMO:\textbf{S} model with the fully minimized reaction coordinate on the trajectory subject to harmonic constraints. $T^\mathrm{full}_\mathrm{rel}$ are for the entire path.
\end{flushleft}
\label{tbl:rhfbarriers}
\end{table}

\newpage
\begin{table}[!ht]
\caption{\bf{Reaction barriers of chorismate mutase calculated using ONIOM.}}
\begin{center}
\begin{tabular}{lrrrrr}
 & $R_R$ & $R_{TS}$ & $R_P$ & $E_{TS-R}$ & $E_{P-R}$ \\ \hline
MP2/6-31G(d) &  -1.83 &   0.13 &   1.56 &  22.24 &  -3.20 \\
MP2/cc-pVDZ &  -1.83 &  -0.36 &   1.56 &  19.75 &  -5.48 \\
MP2/cc-pVTZ &  -1.83 &   0.13 &   1.56 &  21.79 &  -1.14 \\
MP2/cc-pVQZ &  -1.83 &   0.13 &   1.56 &  21.68 &  -0.82 \\
B3LYP/6-31G(d) &  -1.83 &   0.13 &   1.39 &  25.19 &   3.81 \\
B3LYP/cc-pVDZ &  -1.83 &   0.13 &   1.39 &  23.81 &   2.58 \\
B3LYP/cc-pVTZ &  -1.83 &   0.13 &   1.56 &  24.62 &   4.36 \\
B3LYP/cc-pVQZ &  -1.83 &   0.13 &   1.56 &  24.66 &   4.16 \\
\end{tabular}\begin{flushleft}
The reaction coordinates in \AA ~for the reactant, transition state and product are $R_R$, $R_{TS}$ and $R_P$, respectively. The barrier height of the transition state $E_\mathrm{TS-R}$ and the overall reaction energy $E_\mathrm{P-R}$ are in kcal/mol.
\end{flushleft}
\end{center}
\label{tbl:oniombarriers}
\end{table}

\newpage
\begin{table}[!ht]
\caption{\bf{Reaction barriers of chorismate mutase calculated using multilayer EFMO and FMO2 calculations.}}
\begin{center}
\begin{tabular}{lrrrrr}
 & $R_R$ & $R_{TS}$ & $R_P$ & $E_{TS-R}$ & $E_{P-R}$ \\ \hline
EFMO-RHF:MP2/6-31G(d) &  -1.64 &  -0.11 &   1.39 &  27.64 &  -4.70 \\
FMO2-RHF:MP2/6-31G(d) &  -1.64 &  -0.11 &   1.88 &  29.22 &  -6.41 \\
\end{tabular}\begin{flushleft}
The reaction coordinates in \AA ~for the reactant, transition state and product are $R_R$, $R_{TS}$ and $R_P$, respectively. The barrier height of the transition state $E_\mathrm{TS-R}$ and the overall reaction energy $E_\mathrm{P-R}$ are in kcal/mol.
\end{flushleft}
\end{center}
\label{tbl:efmomp2barrier}
\end{table}

\newpage
\section*{Supporting Information}
\begin{flushleft}
\textbf{Figure S1.} Reaction barrier calculated at the MP2/cc-pVDZ:EFMO-RHF/6-31G(d) level of theory for EFMO:\textbf{S} using $R_\mathrm{resdim}=1.5$ and FDD (modfd=3). This snapshot shows the effect of not having enough flexibility in the active region around the substrate.
\end{flushleft}

\begin{flushleft}
\textbf{Figure S2.} Two different starting geometries with chorismate and Phe57 shown as sticks from the MD simulation. A) shows a configuration which results in a successful reaction path and B) a configuration which results in an unsuccessfull reaction path (see Figure~S1). The position of Phe57 coupled with a placement in the buffer region ($b$) makes it unable to move to accomodate the conversion of chorismate to prephenate.
\end{flushleft}

\begin{flushleft}
{\color{black}\textbf{Table S1.} Complete listing of all residues in the protein model (PDB: 2CHT) along with their protonation state after being protonated using the PDB2PQR tool.}
\end{flushleft}

\begin{flushleft}
\textbf{Table S2.} Reaction barriers of chorismate mutase calculated using ONIOM for EFMO:\textbf{S} using $R_\mathrm{resdim}=1.5$ and FD (modfd=1).
\end{flushleft}

\begin{flushleft}
\textbf{Table S3.} Reaction barriers of chorismate mutase calculated using ONIOM for EFMO:\textbf{S} using $R_\mathrm{resdim}=2.0$ and FDD (modfd=3).
\end{flushleft}

\begin{flushleft}
\textbf{Table S4.} Reaction barriers of chorismate mutase calculated using ONIOM for EFMO:\textbf{L} using $R_\mathrm{resdim}=1.5$ and FDD (modfd=3).
\end{flushleft}

\begin{flushleft}
\textbf{Table S5.} Reaction barriers of chorismate mutase calculated using ONIOM for FMO2:\textbf{S} using $R_\mathrm{resdim}=1.5$ and FDD (modfd=3).
\end{flushleft}


\end{document}